# Enabling Lower-Power Charge-Domain Nonvolatile In-Memory Computing with Ferroelectric FETs

Guodong Yin, *Student Member, IEEE,* Yi Cai, Juejian Wu, *Student Member, IEEE,* Zhengyang Duan, Zhenhua Zhu, *Student Member, IEEE,* Yongpan Liu, *Senior Member, IEEE,* Yu Wang, *Senior Member, IEEE,* Huazhong Yang, *Fellow, IEEE,* and Xueqing Li, *Senior Member, IEEE*

*Abstract*—Compute-in-memory (CiM) is a promising approach to alleviating the memory wall problem for domain-specific applications. Compared to current-domain CiM solutions, charge-domain CiM shows the opportunity for higher energy efficiency and resistance to device variations. However, the area occupation and standby leakage power of existing SRAM-based charge-domain CiM (CD-CiM) are high. This paper proposes the first concept and analysis of CD-CiM using nonvolatile memory (NVM) devices. The design implementation and performance evaluation are based on a proposed 2-transistor-1-capacitor (2T1C) CiM macro using ferroelectric field-effect-transistors (FeFETs), which is free from leakage power and much denser than the SRAM solution. With the supply voltage between 0.45V and 0.90V, operating frequency between 100MHz to 1.0GHz, binary neural network application simulations show over 47%, 60%, and 64% energy consumption reduction from existing SRAM-based CD-CiM, SRAM-based current-domain CiM, and RRAM-based current-domain CiM, respectively. For classifications in MNIST and CIFAR-10 data sets, the proposed FeFET-based CD-CiM achieves an accuracy over 95% and 80%, respectively.

## I. INTRODUCTION

THE computing capability and energy efficiency of modern computers based on the von Neumann architecture are hindered by the data movement between the memory component and the processing units, known as the "memory wall" problem [1]. This problem has deteriorated with the advent of the big-data era. To tackle this challenge, recent attempts of computing in the memory (CiM) have become intriguing by reducing the data transfer activities [3][4].

As the conventional memories were not designed for the CiM purpose, a key CiM enabler is to facilitate the memory component with a computable circuit structure and/or a flexible interface, under the constraints of cost, power consumption, scalability, and reliability. From the application perspective, the data-intensive convolutional neural network (CNN) acceleration is of particular interest because of the data formality in computing parallelism and simplicity [2]. Recent exploration works ranging from devices and circuits to architectures and algorithms have indicated the benefits of such a co-design [3]-[5]. These existing CiM methodologies could be roughly classified in two

Manuscript received Apr. 1, 2020. Revised July 18, 2020, Nov. 19, 2020. Accepted Jan. 2, 2021. This work is supported in part by NSFC (#61874066, #61720106013), in part by The National Key Research and Development Program of China (2019YFA0706100, 2018YFA0701500), in part by Key Laboratory of Artificial Intelligence, Ministry of Education, and in part by The Beijing Innovation Center for Future Chips. Corresponding author: X. Li.

All authors are with BNRist, The Department of Electronic Engineering, Tsinghua University, Beijing 100084, China. Email: {ygd20, caiy17, wujj19, duanzy18, zhuzhenh18}@mails.tsinghua.edu.cn, {ypliu, yu-wang, yanghz, xueqingli}@tsinghua.edu.cn.

Digital Object Identifier XXXX/XXXX.XXXX.

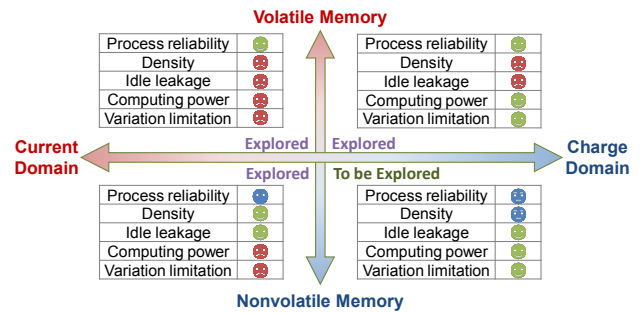

Fig.1. Extending CiM to the NVM-based charge-domain quadrant.

dimensions in Fig. 1: (i) memory devices being volatile or nonvolatile, and (ii) computing and sensing methods being in a current mode or a voltage/charge mode [3]-[6], [28]. Compared with the conventional volatile CMOS solution, NVM has potentially higher density and the intrinsic zero standby power. Compared with the current-mode computing and interfacing, a voltage-mode charge-domain CiM (CD-CiM) consumes only dynamic power, which is appealing for low-power applications. More importantly, the capacitor-based CD-CiM may provide higher immunity to PVT variations, including the on-state current mismatch, which is challenging to handle for both MOSFET and RRAM [5].

Therefore, it is highly motivated and timely to start the adventure of NVM-based CD-CiM in the bottom-right quadrant of Fig. 1 to enable the combined advantages of NVM and charge-domain computing for denser, more reliable, and lower-power solutions. Previously, this was challenging due to the low on/off ratio of MRAM (typically ~2) and RRAM (typically $10^2$-$10^3$ [22]), as to be discussed subsequently.

Today, the emerging of the nonvolatile ferroelectric field-effect-transistor (FeFET) with a highly-scalable CMOS-compatible process and an ultra-high on/off ratio ($>10^6$) has articulated a promising design space for a new CD-CiM paradigm. Besides, the other FeFET features, such as DC-power-free write, separate read and write ports, and the compact integration of NVM and transistor also contribute to more design flexibilities in the device-circuit co-design.

This work proposes the concept, analysis, and design of NVM-based CD-CiM, and exploits FeFETs for the circuit implementation. Evaluation against SRAM and RRAM solutions suggests significantly improved trade-offs between density, performance, and power consumption. Itemized contributions include:

- A FeFET-based 2-transistor-1-capacitor (2T1C) compact CiM cell that supports charge-domain DC-power-free XNOR operations, along with the analysis on the cell-level comparison with other NVM technologies;
- A CD-CiM macro array based on the proposed 2T1C cell for low-power, parallel, and reliable multiply-and-accumulate (MAC) operations in binary neural network (BNN) applications;

- Evaluations of the proposed charge-domain CiM, including (i) 2T1C cell-level CiM circuit performance, (ii) array-level analysis on energy savings of the computing operation and the impact of major array-level variations, and (iii) variation-aware classification accuracy in MNIST and CIFAR-10 datasets.

## II. PROPOSED 2T1C CHARGE-DOMAIN CiM

This section introduces BNN and FeFET background briefly and presents the proposed FeFET-based 2T1C XNOR cell for the MAC CiM computing methodology of BNN applications.

### A. Binary CNN

Convolutional neural networks (CNNs) may achieve high accuracy in computer vision applications, such as image classification and face detection. In CNNs, the multiply-and-accumulate (MAC) is a critical operation during inference and consumes major power [5]. The trained neural networks with binary weights and activations, i.e., +1 and -1, drastically simplify multiplications to be atomic XNOR operations [8]. While BNN has shown its success in smaller data sets like MNIST, recent works have extended the use of low-resolution quantization in larger data sets like ImageNet, showing improved accuracy and much lower costs with algorithm optimizations [8][24].

Binary MAC operations can be performed in three steps: (i) perform XNOR logic in atomic cells; (ii) accumulate the results; (iii) restore the binary value. These operations deliver the corresponding matrix multiplication functions in the convolutional/fully-connected layer and nonlinear functions in the neuron layer. Below, (1) shows the details of the batch normalization and the activation function ρ [9], and (2) shows the particular case for binary batch normalization and the activation function by a *sign* comparison between the MAC result and a reference voltage:

$$OUT_i = \rho(\gamma_i \frac{PA_i - \mu_i}{\sigma_i^2} + \beta_i) \quad (1)$$

$$OUT_i = sign(IN_i - \alpha_i) \quad (2)$$

where $PA_i$ is the pre-activation tensor, $OUT_i$ is the output tensor, $\mu_i$ is the mean of $PA_i$, $\sigma_i$ is the standard deviation of $PA_i$, and $\gamma_i$, $\beta_i$, $\alpha_i$ are batch normalization parameters. Compared to 32-bit CNNs, binary CNNs need only 1/32 memory and also much fewer data accesses, leading to drastic power savings.

XNOR circuits could be implemented in the current mode (current domain) or the voltage mode (charge domain) [3][5]. In the current-mode design, the XNOR results are calculated in custom designed XNOR memory cells based on the Kirchhoff's current law. By identifying the amount of the output current, the sense amplifiers (SA) could tell from different XNOR input scenarios. Due to the DC-power consumed by computing and sensing, the current-mode XNOR operation may not fit well in power-sensitive applications [3][5]. This is more challenging with device variations in the output currents.

In contrast, in the charge domain, XNOR and MAC operations could be implemented more efficiently with the charge conservation law. Fig. 2 illustrates an example using the SRAM array. Each SRAM cell is accompanied by a local capacitor. It performs XNOR operations between the SRAM state and the input pair IA/IAb. Each single XNOR operation result is reflected as the charge stored at the local capacitor near each SRAM in Fig. 2. The charge of the three capacitors is then collected with the source line ScL for a summation of the XNOR results. As the SRAM states are directly linked to the supply voltage VDD or the ground voltage GND (thanks to the high on/off $I_{DS}$ ratio of MOSFETs), the major source that determines the MAC computation accuracy is the mismatch between the capacitors rather than the MOSFET $V_{TH}$ variations. With a proper VDD, this significantly improves the immunity to MOSFET on-state drain current variations (as compared with the current-mode sensing of summed currents). In addition, the operation does not occur with static currents, leading to significant power savings.

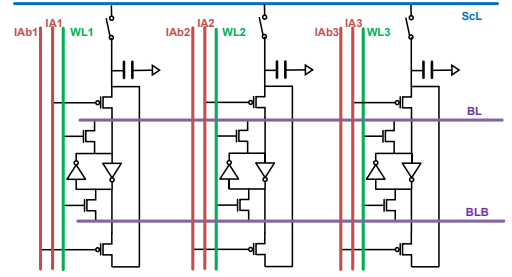

Fig. 2. Existing SRAM charge-domain XNOR and MAC (rotated view) [5].

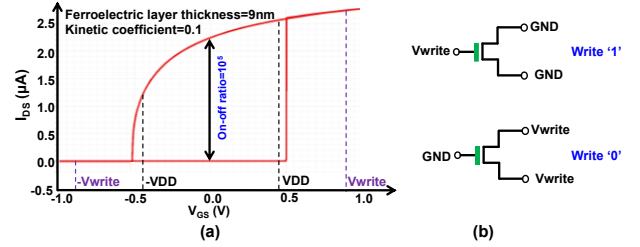

Fig. 3. (a) FeFET $I_{DS}$-$V_{GS}$ curve; (b) write methods.

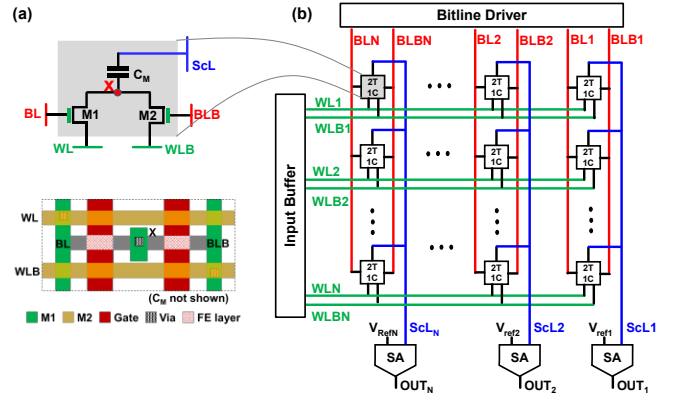

Fig. 4. Proposed FeFET-based CD-CiM: (a) cell; (b) array structure.

The main drawbacks of the SRAM-based charge-domain XNOR and MAC operations are (i) low density due to large SRAM cells, and (ii) idle-state static leakage current. As to be shown subsequently, the proposed new design with FeFETs solves these problems elegantly.

### B. The FeFET Device Basics

An FeFET is essentially a MOSFET with a ferroelectric layer embedded at the gate [13]. The polarization of this extra layer brings a knob to tune and keep a nonvolatile $V_{TH}$, which further leads to a tunable state of the drain-source current $I_{DS}$. Fig. 3. shows the FeFET $I_{DS}$-$V_{GS}$ curve with two states and adopted parameters of the model used in the simulation. Detailed device operating mechanisms have been reported in prior works [12][13][26]. Generally, to reduce (or increase) $V_{TH}$ of an n-type FeFET, a positive $V_{write}$ (or negative $-V_{write}$) voltage pulse could be applied to the FeFET gate. $V_{write}$ should be sufficiently high to trigger partial or full polarization switching. A negative gate voltage may be practiced with a negative $V_{GS}$ to avoid the use of a negative supply. While sensing the FeFET $V_{TH}$ state, the gate should be biased at a moderate voltage below $V_{write}$ to prevent disturbance. Recent use of hafnium-based materials makes FeFETs well scalable [14][15]. With the amplification of the embedded MOSFET, $I_{DS}$ could exhibit an ultra-high range beyond $10^6$, which is particularly preferred for computing in large memory arrays [17][29]. FeFETs also own a moderate endurance (up to $10^{12}$ [18]), a moderate operating voltage range (low to 1.5V [13]), and speed (up to ns [19]).

Notably, recent reports show that highly-scaled FeFETs maintains the high on/off ratio, but may also suffer from large $I_{DS}$ variations. This limits the accuracy of current-sensing-based CiM [20]. Therefore, innovations that exploit the ultra-high on/off ratio rather than the absolute $I_{DS}$ is more preferred for computing purposes. This is the contribution of this work when compared with existing FeFET-based current-mode solutions.

*C. Proposed 2T1C CD-CiM macro*

*Cell and array structures*. Fig. 4 (a) shows the proposed 2T1C cell. It consists of two n-type FeFETs (M1 and M2), and one capacitor $C_M$. WL and WLB are wordlines. BL, BLB, and ScL are bitlines. The cell can store bits as FeFET states: '1' for positive polarization state (negative $V_{TH}$) and '0' for negative polarization state (positive $V_{TH}$). In the cell, the two FeFETs store one '0' and one '1', similar to SRAM. Fig. 4 (b) shows the proposed FeFET-based CD-CiM array macro implementation based on the 2T1C cells. Multiple rows and columns can be activated simultaneously to compute in parallel.

*Cell and array write operations*. TABLE I shows the write setup, with an example of writing '1' to M1 and '0' to M2 in one cell. It has two phases (Phase 1 and Phase 2), similar to the methods in [12][20]. In TABLE I, $V_{BL}/V_{BLB}$ is set to $V_{write}$/GND to write '1' to M1 and '0' to M2. In Phase 1, $V_{WL}$ and $V_{WLB}$ are connected to GND. In Phase 2, $V_{WL}$ and $V_{WLB}$ are driven concurrently to $V_{write}$ for a period of time. An effective write of '0' and '1' occurs with $V_{GS} = -V_{write}$ and $V_{GS} = V_{write}$, respectively. The write operations of one cell could be easily extended to an array. WL/WLB of the selected row are connected to GND in Phase 1 and $V_{write}$ in Phase 2. WL/WLB of unselected rows are set to $V_{write}/2$, and BL/BLB is set to $V_{write}$ to write '1' for the selected FeFET, or GND to write '0' for the selected FeFET. For unselected rows, $|V_{GS}| = V_{write}/2$ is maintained to avoid state disturb.

*Cell XNOR operation*. Fig. 5 shows the XNOR logic in one cell. Initially, all bitlines and wordlines are set to GND. Then, ScL is left floating at GND. As mentioned above, either M1 or M2 has low resistance, so the internal node voltage $V_X$ is GND. Next, $V_{WL}/V_{WLB}$ is set to VDD/GND and GND/VDD for an input pair of '1/0' and '0/1', respectively. Note that VDD is set lower than $V_{write}$ to avoid FeFET state disturbance. With the complementary '0' and '1' storage within each cell, when WL or WLB biased at VDD is connected to an on-state FeFET, $V_X$ will be pulled up to VDD. Otherwise, $V_X$ remains GND. As ScL is floating, the change of $V_X$ is linearly delivered to the top plate of $C_M$, i.e., ScL. Fig. 5 has illustrated $V_{ScL}$ in two input scenarios (without considering the impact of parasitics capacitance): equal to $V_{WL}$=VDD in Fig.5(a) for output '1', and equal to $V_{WLB}$=GND in Fig. 5(b) for output '0'. Fig. 6 shows a snapshot of transient simulation results, where both the voltage of bitlines and wordlines and FeFET polarization are included for XNOR outputs of '1' and '0'. Operating non-idealities will be analyzed in Section III.

*Array MAC operation*. Fig. 7 shows the array MAC operation with the shared ScL, BL/BLB, and WL/WLB between unary XNOR CiM cells. At the array-level MAC operation, the output $V_{ScL}$ is driven by multiple XNOR cell outputs in each column. Note that cells in the same column have the same inputs with shared BL and BLB. Therefore, $V_{ScL}$ will be lifted linearly as a function of the number of pulled-up cells. If M cells out of a total of N XNOR cells deliver '1', $V_{ScL}$ is shifted from GND to VDD*M/N.

Given a mapping scheme between -1/+1 and GND/VDD, the pre-trained weights are stored as FeFET states in the array, and the input signals are set through the WL/WLB lines. For the convolutional layer, the weights of the same filter are stored in the same column. For the fully connected layers, the weights are loaded similarly. Every column performs MAC operations, then the nonlinear activation and the binary batch normalization operations are performed with the outputs presented at ScL. Wordlines (WL and WLB) and bitlines (BL and BLB) could be set to GND to make corresponding rows and columns inactive so as to support a smaller network. Large networks may also be supported by matrix splitting with several smaller CiM macros, as discussed in [23].

TABLE I: WRITE OPERATION CONFIGURATION

| Target example | Phases | $V_{ScL}$ | $V_{BL}$ | $V_{BLB}$ | $V_{WL}$ | $V_{WLB}$ |
|---|---|---|---|---|---|---|
| M1: '1'; M2: '0' | Phase1 | GND | $V_{write}$ | GND | GND | GND |
|  | Phase2 |  |  |  | $V_{write}$ | $V_{write}$ |

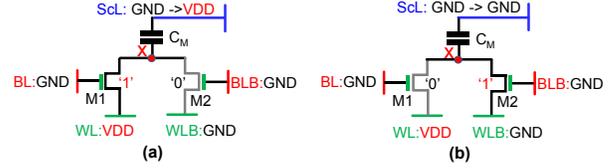

Fig. 5. Proposed XNOR logic of one cell: (a) '1' output; (b) '0' output.

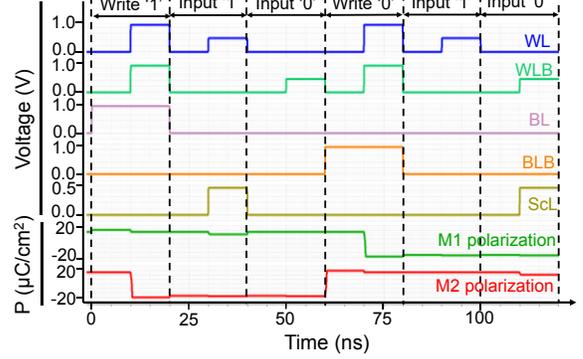

Fig. 6. Transient waveforms of the 2T1C cell (settings see III.A).

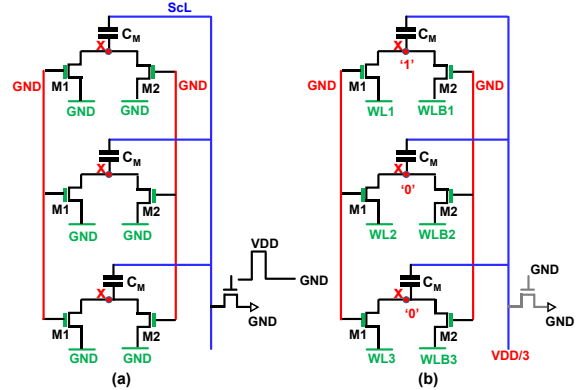

Fig. 7. Proposed 2-phase MAC operation (3 cells in a column example): (a) discharge capacitors and keep ScL floating; (b) XNOR and accumulate.

## III. EVALUATION AND DISCUSSION

This section evaluates the energy, area, and accuracy performance of the proposed 2T1C MAC CD-CiM macro against other existing techniques, in the presence of non-idealities.

*A. Benchmark Settings*

The 2T1C cells are simulated using the calibrated FeFET SPICE model in [21], as shown in Fig. 3. This model has been adopted for prior circuit works. All MOSFETs are the 10nm PTM models [27]. In the benchmarking, $C_M$ is 1.2fF for all charge-domain solutions, adopted from [5]. We use $R_{ON}$ = 10KΩ and $R_{OFF}$ = 1MΩ for RRAMs. The array size is set to 128x128 as a typical case. RC parasitic parameters are from [23]. As shown in recent reports, the SA may consume a significant portion of energy. For fair comparisons, the adopted SAs are based on those in [7].

## B. Energy And Latency Evaluation

### 1) Theoretical Analysis

In CD-CiM, most energy is consumed by capacitance charging. Fig. 8 shows the capacitor network model in the MAC operation of a column. During the charging process, the right-side plate of $C_M$ in Fig. 8 (a) is kept floating and the left-side plate, i.e., node X, is clamped to VDD or GND according to the XNOR result. The equivalent charging capacitor $C_{EQ}$ of a column could be calculated:

$$C_{EQ} = M \times (128 - M) \times C_M / 128, \quad (3)$$

where $M$ is the number of cells whose XNOR result is '1'. As observed, there is no charging load when all cells are delivering XNOR results of '0' or '1', as ScL is floating. The maximum charging load occurs when half cells are delivering '1' (and the other half delivering '0').

Differently, in the SRAM-based CD-CiM design [5], $C_M$ is charged when the XNOR result is '1', and is not charged when it is '0'. Fig. 8(c) compares the equivalent capacitor load, in which the proposed design at $p_1$=0.5 is only half of the SRAM-based design. Here, $p_1$ denotes the percentage of XNOR cells in a column that produce '1'. On average, $C_{EQ}$ of 2T1C design is only 33% of the SRAM-based design. Also, the 2T1C design consumes no idle power with the FeFET non-volatility, which also outperforms the SRAM-based design.

For current-domain CiM solutions, the energy is consumed while settling down and sensing the bitline currents, along with maintaining the reference currents. Although a latch-style dynamic current-SA could be used, there still a trade-off between the current amplitude, supply and the operating frequency to minimize the energy.

### 2) Experimental Simulation

To comply with the FeFET model, the supply voltage is set to 0.45V to avoid FeFET state disturbance. For other designs, an extra 0.90V supply is provided to investigate more options. In the evaluation, we set a clock cycle as the time window for each bitline and wordline controls, including the precharging and clamping, and a clock cycle for sensing and latching the outputs. Evaluations are done at 100MHz and 1.0GHz, each with custom optimizations, e.g. low or high $V_{TH}$ options.

The comparison of energy consumption with related works is shown in Fig. 9. This work achieves the highest energy efficiency. Compared with the current-domain solutions, the minimum improvement is 2.5x at 1.0GHz, and up to 24x at 100MHz in which more time is spent on bitline settling-down. Practically, the operating frequency could be limited by the influence of the PVT variations in current-sensing CiM.

Compared with the SRAM-based CD-CiM, the energy efficiency improvement is 1.9x at 0.45V and 7.8x at 0.90V, which confirms the theoretical analysis above. CD-CiM evaluation results are not sensitive to the frequency unless one fails to reach the operation speed. For example, the SRAM-based CD-CiM fails to reach 1GHz at 0.45V.

## C. Precision Analysis

### 1) Theoretical Analysis and Array-Level Simulation

The energy evaluation above has assumed no variation impact. However, in the current-mode sensing, variations could be playing a key role as the variations of currents directly affect the summed result. For the proposed CD-CiM, it is also important to investigate how the variations of FeFET and $C_M$ affect the overall computing accuracy.

Intuitively, as FeFETs have a very large on-off ratio, the drain-source leakage current $I_{OFF}$ is negligible when compared with the on-state current $I_{ON}$. Therefore, the internal node $X$ in each cell is well set at GND or VDD. Further, the non-ideality of the computation based on the charge re-distribution is determined by the $C_M$ capacitor mismatch, which affects the amount of charge re-distribution at the output ScL. Theoretically, the MAC result from (3) is reshaped as

$$V_{MAC} = V_{ScL} = \frac{1}{\sum_{i=1}^{N} C_i} \sum_{i=1}^{N} V_{Xi} \times C_i$$
$$= \frac{VDD}{\sum_{i=1}^{N} C_i} \times \left( \sum_{i=1}^{M} \frac{C_i \times R_{OFFi}}{R_{ONi} + R_{OFFi}} + \sum_{i=M+1}^{N} \frac{C_i \times R_{ONi}}{R_{ONi} + R_{OFFi}} \right). \quad (4)$$

where $C_i$ is the capacitor of the $i^{th}$ cell, $V_{Xi}$ is $V_X$ of the $i^{th}$ cell, $R_{ONi}$ is the on-state FeFET drain-source resistance of the $i^{th}$ cell, $R_{OFFi}$ is the off-state FeFET drain-source resistance of the $i^{th}$ cell, $N$ is the total number of cells in the MAC operation, and $M$ is the number of cells whose XNOR result is '1'.

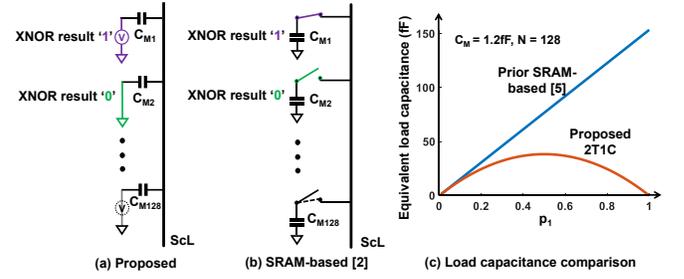

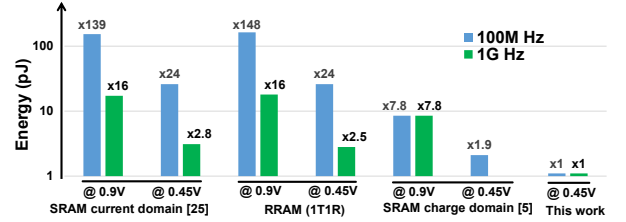

Fig. 8. ScL charging capacitance: (a) Model of proposed 2T1C design; (b) Model of prior SRAM-based design; (c) Comparison as a function of $p_1$.

Fig. 9. MAC operations comparisons between different works in an array.

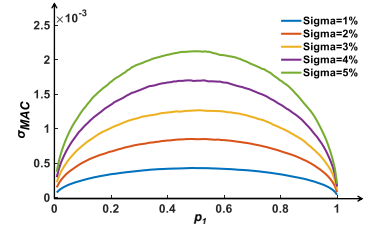

Fig. 10. Normalized standard deviation $\sigma_{MAC}$ vs $p_1$.

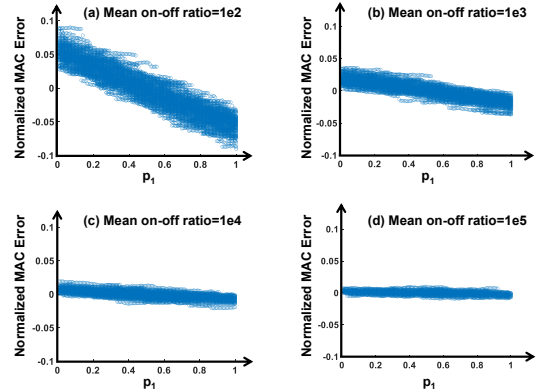

Fig. 11. Normalized MAC error vs $p_1$ with $\sigma_C$=5% and different on-off ratios.

In the analysis, $N$ is set to 128; $C_M$ is modeled as a Gaussian distribution with a mean value of 1.2fF. With the normalized standard $C_M$ deviation $\sigma_c$ between 1% - 5% and the on-off ratio set infinite, Fig. 10 shows the normalized standard deviation of $V_{MAC}$. Considering all corners, the worst $\sigma_{MAC}$ occurs at {$p_1$=0.5, $\sigma_c$=5%} and is below 0.25%. This indicates a much smaller impact than the direct current summing errors caused by typical RRAM or MOSFET $I_{ON}$ variations.

Fig. 11 shows normalized MAC errors with a different on-off ratio, in which $R_{ON}$ and $R_{OFF}$ are logarithmic Gaussian random variables with a normalized standard deviation of 15% and $\sigma_C$=5%. When the on-off ratio is over $10^5$, the normalized MAC error has a chance of ~99.2% to be below that caused by flipping an XNOR cell. In contrast, with an on-off ratio around $10^2$, which is a typical value for RRAM, the accumulated error could be so significant that the average normalized MAC error could be as high as 5%. This finding actually answers the fundamental question: why is it challenging to explore CD-CiM using RRAM and MTJ in the fourth quadrat in Fig. 1?

*2) Application Simulation*

The proposed FeFET-based CD-CiM macro is evaluated for classification applications while considering the impact of $C_M$ variations. The Pytorch framework is used to build the binary LeNet on the MNIST test set [11] and the binary NIN [10] on the CIFAR-10 test set. To evaluate the effect of the non-idealities of the core array, it is assumed that peripherals, such as the reference voltage generator, the SAs, and the quantization blocks, do not lower the overall accuracy.

Fig. 12 scatters the classification accuracy as a function of $\sigma_C$. Ideally, XNOR-Net in [8] achieves ~ 99.0% classification accuracy on the MNIST test set and 85.6% classification accuracy on the CIFAR-10 test set. As shown in Fig. 12, as long as the capacitor mismatch is within a reasonable range of 20% for CIFAR-10 and 30% for MNIST, the classification accuracy is almost uncontaminated. In practical designs, this matching requirement could be used to guide the $C_M$ design given a specific technology for the optimized trade-off between the target accuracy, power consumption, and the layout area.

*D. Area*

The proposed FeFET-based cell consists of two transistors and one capacitor. Because the capacitor can be placed on top of the transistors, the area overhead of the capacitor is significantly reduced. In contrast, the SRAM-based CD-CiM cell needs one capacitor and a total of 9 transistors, including 8 for XNOR cell and one extra transistor to connect the cell to ScL for accumulation in a MAC [5]. In addition, the 2 transistors in the proposed design are both n-type and could be placed in a more compact layout than SRAM transistors.

## IV. CONCLUSIONS

This paper has presented the concept and design of an NVM-based charge-domain computing-in-memory approach. A 2T1C XNOR CiM cell is proposed based on FeFET, a nonvolatile CMOS-compatible NVM device with an ultra-high on/off ratio. The array implementation for MAC CiM macro based on the proposed cell is presented and evaluated. Comparisons show higher density and lower power than prior current-domain and charge-domain CiM designs. Circuit and application evaluations have shown the potential of improving the performance and energy efficiency of BNN accelerators while achieving high accuracy.

## Acknowledgment

The authors would like to thank Prof. Nan Sun for helpful discussions, and Prof. Sumeet Gupta and Kai Ni for model support.

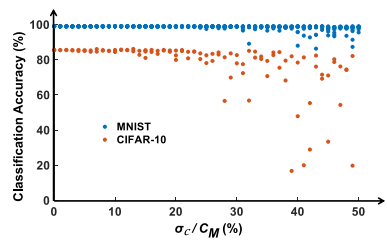

Fig. 12. Impact of $C_M$ mismatch on classification.